\documentclass[11pt,fleqn,twoside]{article}
\usepackage{latexsym}
\makeatletter
\newcommand{\prava}{\footnotesize\it
\begin{flushright}
\begin{minipage}{18cm}
Copyright \copyright 1998 by S.G. Bindu and V.C. Kuriakose
\end{minipage}
\end{flushright}}

\newcommand{\name}[1]{\begin{flushleft}
                       \LARGE \bf #1
                       \end{flushleft}\vspace{-3mm}}

\newcommand{\Author}[1]{\begin{flushleft}
                       \it #1 \end{flushleft}}

\newcommand{\Adress}[1]{\begin{flushleft}
                       \it #1 \end{flushleft}}

\newcommand{\Date}[1]{\begin{flushleft}
                      \small  \it #1 \end{flushleft}}

\newcommand{\ehkol}{Author \ name}
\newcommand{\ohkol}{Article \ name}
\renewcommand{\@evenhead}{
\hspace*{-3pt}\raisebox{-15pt}[\headheight][0pt]{\vbox{\hbox to \textwidth
{\thepage \hfil \ehkol}\vskip4pt \hrule}}}
\renewcommand{\@oddhead}{
\hspace*{-3pt}\raisebox{-15pt}[\headheight][0pt]{\vbox{\hbox to \textwidth
{\ohkol \hfil \thepage}\vskip4pt\hrule}}}
\renewcommand{\@evenfoot}{}
\renewcommand{\@oddfoot}{}

\newcommand{\be}{\begin{equation}}
\newcommand{\ee}{\end{equation}}
\newcommand{\ba}{\hspace*{-5pt}\begin{array}}
\newcommand{\ea}{\end{array}}

\newcommand{\ds}{\displaystyle}
\makeatother

\begin{document}
\setcounter{page}{149}
\input{psfig}
\thispagestyle{empty}

\renewcommand{\ehkol}{S.G. Bindu and V.C. Kuriakose}
\renewcommand{\ohkol}{Nonlinear Wave Propagation Through Cold Plasma}

\begin{flushleft}
\footnotesize \sf
Journal of Nonlinear Mathematical Physics \qquad 1998, V.5, N~2,
\pageref{bindu-fp}--\pageref{bindu-lp}. \hfill {\sc  Letter}
\end{flushleft}

\vspace{-5mm}

\renewcommand{\footnoterule}{}
{\renewcommand{\thefootnote}{}
\footnote{\prava}}

\name{Nonlinear Wave Propagation  Through \\ Cold Plasma}\label{bindu-fp}

\Author{S.G. BINDU  and V.C. KURIAKOSE}

\Adress{ Department of Physics, Cochin University of Science and
Technology, \\
Cochin-682022, India}

\Date{Received Jule 15, 1997; Accepted January 15, 1998}

\begin{abstract}
\noindent
Electromagnetic wave propagation through cold collision free
plasma is studied using the nonlinear perturbation method. It is found
that the equations can be reduced to the  modified Kortweg-de Vries equation.
\end{abstract}


\section{Introduction}

An exciting  and extremely active area of research investigation  during
the past years has been the study of solitons and the related issue of
the construction of solutions to a wide class of nonlinear equations. The
concept of solitons has now become umbiquitous in modern nonlinear
science and indeed can be found in various branches of physics. In
nonlinear wave propagation through continuous media, steepening of waves
arises due to nonlinearities which is balanced by dissipative or dispersive
effects. Exciting and important discoveries were
made in the nonlinear dynamics of dissipative and conservative systems.
There are different methods to study nonlinear systems. The reductive
perturbation method for the propagation of a slow modulation of
a quasimonochromatic wave was first established by Taniuti and Washimi
for the whistler wave in a cold plasma.
This method was generalised to a wide class of nonlinear wave systems by
Taniuty and Yajima. Kakutani and Ono [1], Kawahara and Taniuti [2] and
Taniuti and Wei [3] have investigated the propagation of hydromagnetic
waves through a cold collision free plasma using this reductive
perturbation method.

In the study of the asymptotic behaviour of nonlinear dispersive waves,
Gardner and Morikava [4] were the first to introduce the scale
transformation
\[
\xi= \varepsilon^{ \alpha} (x-vt), \qquad
\tau= \varepsilon^{ \beta} t.
\]
This scale transformation is called the Gardner-Morikawa [4]
transformation.  They combined this transformation with a perturbation
expansion of the dependent variables so as to describe the nonliear
asymptotic behaviour and in the process they arrived at the Kortweg
de-Vries [KdV] equation [5] which is a single tractable equation
describing the asymptotic behaviour of a wave. This method has established
a systematic way for the reduction of a fairly general nonlinear systems
to a single tractable nonlinear equation describing the far field
behaviour. The reductive perturbation method was first established for
the long wave approximation and then for the wave modulation problems.

In the present work we study the propagation of electromagnetic
waves through a cold collision free plasma by using a nonlinear reductive
perturbation method.  It is found that to the lowest order of
perturbation the system of equations can be reduced to the modified
Kortweg-de Vries equation (mKdV) [6]. In the case of steady state
propagation this equation can be integrated to give
a solution in terms of hyperbolic functions which exhibit solitary wave
nature.

\section{Formulation of the problem}

When electromagnetic waves pass through a medium , the system gets
perturbed. Since electrons are much lighter than  ions, electrons respond
much more rapidly to the fields and ion motion can be neglected. In the
equation of momentum for cold plasma, no pressure term is present. Basic
equations relevant to the present problem are the equations of motion of
electron and the Maxwell's equations. Here we are interested only in the
electronic motion.  To obtain a single equation which incorporates weak
nonlinear and weak dispersive effects, we employ the expansions of the
dependent variables similar to that introduced by  Nakata  [7].

The equation of motion of an electron in an electromagnetic field is
\begin{equation} \label{2}
\frac{ d \vec v}{ d t}=-\frac em \left[ \vec E+ \vec v\times \vec B\right].
\end{equation}
Taking the leading order terms we get,
\[
\frac{\partial \vec v}{\partial t}=-\frac{e}{m}\left[
\vec E+ \vec v\times \vec B\right].
\]
For convenience we take the displacement vector field $ \vec S$, which
describes the direction and distance that the plasma has moved from the
equilibrium. That is,
\[
\vec v=\frac{\partial \vec S}{\partial t}+(\vec v \cdot\nabla) \vec S.
\]
Therefore  Eq.(\ref{2}) can be written as
\begin{equation} \label{5}
\frac{\partial^2 \vec S}{\partial t^2}= -\frac{e}{m}\left[\vec
E+\left(\frac{\partial \vec S}{\partial t}\times \vec B\right)\right].
\end{equation}

From Maxwell's equation
\[
\nabla\times \vec E=-\frac{\partial \vec B}{\partial t},
\]
and  writing
\[
E =E_0\exp{i(kx-\omega t)}
\]
  we obtain:
\begin{equation} \label{8}
kE=\omega B.
\end{equation}
Substituting   Eq.(\ref{8}) in  Eq.(\ref{5})    we can write:
\begin{equation} \label{9}
\frac{\partial^2 \vec S}{\partial t^2}=
\vec E+\left[\frac{\partial \vec S}{\partial t}\times \vec E\right]W,
\end{equation}
where      $\ds W= \frac{k}{\omega} $.
Physically each electron is acted upon by an electric field that is
parallel to the magnetic field so that there is no perpendicular
component of motion that could be affected by the Lorentz force, that is
$ \vec v\times \vec B = 0 $~[8].
Then Eq.(\ref{2})   can be written as
\[
m\frac{\partial \vec v}{\partial t} =-e \vec E.
\]
 Equation (\ref{9}) then  becomes
\[
\frac{\partial^2 \vec S}{\partial t^2}=
\frac{\partial \vec v}{\partial t}+\left(\frac{\partial \vec S}
{\partial t}\times \vec E\right)W
\]
which can be put as:
\begin{equation} \label{12}
\frac{\partial \vec S}{\partial t}=\vec v+(\vec S\times\vec E)W.
\end{equation}

From Maxwell's equations, we have
\begin{equation} \label{13}
   \vec B = \mu_0 \vec H, \qquad
\nabla \times \vec E=-\frac{\partial \vec B}{\partial t},
\end{equation}
\begin{equation} \label{14}
\nabla\times \vec H=\vec J+\varepsilon_{0}\frac{\partial \vec E}
{\partial t}.
\end{equation}
Taking the time derivative of Eq.(\ref{14})  we get,
\begin{equation} \label{15}
\nabla\times\nabla\times \vec E=1/c^2\frac{\partial^2 \vec E}{\partial
t^2}+ \mu_0 \frac{\partial \vec v}{\partial t}.
\end{equation}
 Equation (\ref{15})  can then be written as
\begin{equation} \label{16}
c^2 \nabla^2 \vec{E}= \frac{\partial^2 \vec{E}}{\partial
t^2}+\frac{\partial^2 S}{\partial t^2}.
\end{equation}
Equations (\ref{12}) and (\ref{16}) are systems of complicated nonlinear
partial differential equations for $ \vec E $ and $\vec S $ describing
electromagnetic wave propagation through plasma.
Let us seek a solution of these equations in the form of a Fourier
expansion in harmonics of the fundamental $ E =e^{i(kx-\omega t)} $ as,
\begin{equation}\label{17}
E =\sum_{n=-\infty}^{+\infty} \vec E^nE^n,
\end{equation}
\begin{equation}\label{18}
S =\sum_{n=-\infty}^{+\infty} \vec S^nE^n.
\end{equation}

Let us now consider  one dimensional plane wave propagating along the
$x$ direction in the Cartesian coordinate system $(x,y,z)$. All the
physical quantities are assumed to be functions of one space coordinate $x$
and time $t$. We now introduce the streching variables
$\xi$  and  $\tau$ as,
\[
\xi = \varepsilon(x-Vt),
\]
\[
\tau = \varepsilon^3t,
\]
where the velocity can be determined by the solvability condition of the
above equations.

$E$ and $S$    satisfy the  following boundary conditions,
\begin{equation} \label{21}
E_{x}^{i} \to 0 \quad \mbox{except} \quad E_{x}^{0}=E_{0}\cos{\theta},
\end{equation}
\begin{equation}\label{22}
E_{y}^{i}\to 0 \quad \mbox{except} \quad E_{y}^{0}=E_{0}\sin{ \theta},
\end{equation}
\begin{equation}\label{23}
E_{z}^{i} \to 0,
\end{equation}
\begin{equation}\label{24}
\mbox{as}\qquad  \xi\to - \infty, \qquad i=  0,1, 2,3,\ldots
\end{equation}
\begin{equation}\label{25}
S _{x}^{i}\to  0  \quad \mbox{except}\quad S_{x}^{0} =S_{0}\cos{\theta},
\end{equation}
\begin{equation}\label{26}
S _{y}^{i}\to 0 \quad \mbox{except}\quad S_{y}^{0}=S_{0}{\sin\theta},
\end{equation}
\begin{equation} \label{27}
S_{z}^{i}\to 0,
\end{equation}
\begin{equation}\label{28}
\mbox{as}\quad  \xi\to - \infty,\qquad i= 0,1,2,3, \ldots
\end{equation}
The  operators   in terms of the streching variables
can be written as
\[
\frac{\partial}{\partial t} = (-\varepsilon v)\frac{\partial}{\partial
\xi}+\varepsilon^2\frac{\partial}{\partial \tau},
\]
\[
\frac{\partial^2}{\partial t^2} =
(v^2\varepsilon^2)\frac{\partial^2}{\partial\xi^2}-2\varepsilon^4
\frac{\partial}{\partial \xi}\frac{\partial}{\partial \tau},
\]
\[
\frac{\partial ^2}{\partial x ^2}=(\varepsilon^2)\frac{\partial}{\partial
\xi^2}.
\]
For  an   appropriate choice of the coordinate system we can write
$E=(E_x,E_y,0)$ and $S=(S_x,S_y,0)$. Expressing the Fourier components of
$E$ and $S$ in powers of a small parameter $\varepsilon $
\[
S^n=\sum_{j=0}^{\infty}\varepsilon^jS_J^n(x,t),
\qquad
E^n=\sum_{j=0}^{\infty}\varepsilon^jE_j^n(x,t).
\]

Before proceeding to the nonlinear problem, it may be instructive to
examine the dispersive relation in the linearized limit. Assuming a
sinusoidal wave $ \exp i(kx-\omega t) $, where $k$, $ \omega $ are
respectively the wave number and the frequency of the wave. Expanding the
above coupled equations,
\[
\left[\frac{\partial}{\partial t}    -in \omega\right] S^n=
\sum_{p+q=n}  {S^p \times E^q},
\]
\[
\left [\frac{\partial ^2}{\partial t^2}+2in \omega
\frac{\partial}{\partial t}-n^2 \omega^2\right] [E_s^n+S_s^n]=
c^2\left[\frac{\partial ^2}{\partial x^2}+2ink \frac{\partial}{\partial
x}-n^2 k^2\right]E_s^n(1- \delta_s,x)
\]
Eq.(\ref{16}) gives the components of $ S_1,_s^n $ as funtions of $
E_1,_s^n $,     $   (s=x,y,z) $.

The determinant of this  system, $ \Delta(n) $ is
\[
\Delta(n)=in\omega\left[-n^2\gamma^2\omega^2+\mu^2s_x^2+\gamma\mu(1+
\alpha)s_t^2\right],
\]
where
\begin{equation}\label{36}
\mu=(1+\alpha\gamma)W, \quad
\gamma=\left(1-\frac{k^2}{\omega^2}\right),\quad
\alpha=\frac{E_0}{S_0}.
\end{equation}

For n=1, $ \Delta(1) $ is zero if $ \omega $ satisfies the dispersion
relation
\[
-\gamma^2\omega^2+\mu^2s_x^2+\gamma\mu(1+\alpha)s_t^2=0.
\]

From    this   we    obtain  $\ds  V=\sqrt{\frac{\alpha}{(1+\alpha)}} c$.
We  assume that $ S_0^0=s $ and $ E_0^0=\alpha s $ are constants and that
\[
M_0^n=H_0^n=0 \qquad \mbox{for}\quad n  \ne 0.
\]
The assumed conditions at infinity are, $ E_j^n, S_j^n \to 0 $ for $ j =
0 $ for all ``$n$'' except for
$(j,\mid {n}\mid )=(1,1),$ where the limit is assumed to be a finite
constant. For $n=1$, $ \Delta (1) = 0 $  for $j=1$.
Under this condition the system has a nontrivial solution. But for
$n= 2,3,4, \ldots$,
$ \Delta n \ne0 $, we have the trivial solution. That is for $j=1$ and
$n > 1$, we get  $ E_1^n=S_1^n=0 $.

For $n = 0$, $ \Delta(0) = 0 $, we can choose $ E_1^0=S_1^0=0 $. This
completes the solution at order $(1,n)$.

For the next order, we can proceed in the same manner.
The system  will have a solution
only if the determinant of the augmented matrix is zero .

Now expanding the  dependent variables as,
\[
S_x=S_0+\varepsilon ^1 S_x^1+\varepsilon^2 S_x^2+\cdots,
\]
\[
S_y=S_y^0+\varepsilon^1 S_y^1+\varepsilon^2 S_y^2+\cdots,
\]
\[
S_z=S_z^0+\varepsilon^1 S_z^1+\varepsilon^2 S_z^2+\cdots,
\]
\[
E_x=E_x^0+\varepsilon^1 E_x^1+\varepsilon^2 E_x^2+\cdots,
\]
\[
E_y=E_y^0+\varepsilon^1 E_y^1+\varepsilon^2 E_y^2+\cdots,
\]
\begin{equation}\label{46}
E_z=E_z^0+\varepsilon^1 E_z^1+\varepsilon^2 E_z^2+\cdots.
\end{equation}
Subtituting these expansions in Eqs.(\ref{12}) and (\ref{17}), then
collecting and solving coefficients of different orders of $ \epsilon ^j
$ for $ n=1 $ with the boundary conditions given by Eqs.(\ref{21}) to
(\ref{26})  we get:\\
at order $ \varepsilon^0 $
\[
S_y^0E_z^0-S_z^0E_y^0=0,
\]
\[
S_z^0E_x^0-S_x^0E_z^0=0,
\]
\[
S_x^0E_y^0-S_y^0E_x^0=0,
\]
\[
\frac{\partial^2 E_x^0}{\partial \xi^2}=0,
\]
\begin{equation}\label{51}
V^2\frac{\partial^2 }{\partial \xi^2} (\gamma E_y^0+S_y^0)=0;
\end{equation}
at order $ \varepsilon^1 $
\[
(1+ \alpha)S_z^0S_x^1=-V\frac{\partial S_y^0}{\partial \xi},
\]
\[
(1+ \alpha)S_y^0S_x^1=-V\frac{\partial S_z^0}{\partial \xi},
\]
\[
(E_x^1+S_x^1)=0,
\]
\begin{equation}\label{55}
(E_y^1- \alpha S_y^1)=0,
\end{equation}
\begin{equation}\label{56}
(E_z^1- \alpha S_z^1)=0;
\end{equation}
at order $ \varepsilon^2 $
\[
V \frac{\partial S_x^1}{\partial \xi}=S_y^0(E_z^2-\alpha
S_z^2)-S_z^0(E_y^2-\alpha S_y^2),
\]
\begin{equation}\label{58}
V \frac{\partial S_y^1}{\partial \xi}=S_z^0(E_x^2-\alpha
S_x^2)-S_x^0(E_z^2-\alpha S_z^2),
\end{equation}
\[
V \frac{\partial S_z^1}{\partial \xi}=S_x^0(E_y^2- \alpha
S_y^2)-S_y^0(E_x^2- \alpha S_x^2)+(1+ \alpha )S_x^1S_y^1,
\]
\begin{equation}\label{60}
(E_x^2+S_x^2)=0,
\end{equation}
\begin{equation}\label{61}
\frac{\partial (E_y^2-\alpha S_y^2)}{\partial \xi}=-2V\frac{(1+\alpha
)^2}{c^2}\frac{\partial S_y^0}{\partial \tau},
\end{equation}
\[
\frac{\partial (E_z^2-\alpha S_z^2)}{\partial \xi}=-2V\frac{(1+\alpha
)^2}{c^2}\frac{\partial S_z^0}{\partial \tau}.
\]
Solving for $ E_y^2$, $S_y^2$ and $E_z^2$, $S_z^2 $, from Eqs.(\ref{55})
and (\ref{56})    we  can get
\begin{equation}\label{63}
(E_y^2- \alpha S_y^2)=\int -2V(1+ \alpha)^2/c^2 \frac{\partial
S_y^0}{\partial \tau}{d \xi}, \quad
(E_z^2- \alpha S_z^2)=\int -2V(1+ \alpha )^2/c^2 \frac{\partial
S_z^0}{\partial \tau}{d \xi}.
\end{equation}
\[
V\frac{\partial S_x^1}{\partial \xi}=S_y^0(E_z^2- \alpha
S_z^2)-S_z^0(E_y^2- \alpha S_y^2).
\]
Substituting for
\[
(E_y^2- \alpha S_y^2)\qquad \mbox{and} \qquad (E_z^2- \alpha S_z^2)
\]
from Eq.(\ref{61})  in Eq.(\ref{58}) we get
\[
S_y^0 \int \frac{-2V(1+ \alpha )^2}{c^2}\frac{\partial}{\partial
\tau}S_z^0{d \xi} -S_Z^0 \int\frac{-2V(1+ \alpha
)^2}{c^2}\frac{\partial}{\partial \tau}S_y^0{d \xi}=V\frac{\partial
S_x^1}{\partial \xi},
\]
\[
S_y^0\int\limits_{- \infty}^{ \xi}\frac{\partial }{\partial \tau}S_Z^0 d{
\xi}-S_z^0 \int\limits_{- \infty}^{ \xi}\frac{\partial}{\partial
\tau}S_y^0 d{ \xi}=-\frac{c^2}{2(1+ \alpha)^2}\frac{\partial
S_x^1}{\partial \xi}.
\]
Now  introducing   two new variables $A$ and $ \theta $ defined by
\[
S_y^0=A \cos {\theta}, \quad S_z^0  = A  \sin { \theta},
\quad A = S_0 \sin { \phi} \quad \theta \to 0 \quad
\mbox{as}\quad     \xi \to  \infty.
\]
Equation (\ref{46}) can be written as
\[
S_x^1=-\frac {V}{(1+ \alpha)}\frac{\partial  \theta}{\partial \xi}.
\]
Now subtituting the value of $ S_x^1 $ and using the new variables
Eq.(\ref{60}) can be written as,
\[
\cos {\theta} \frac{\partial}{\partial \tau} \int\limits_{-
\infty}^{\xi}\sin{ \theta}{d \xi}-
\sin {\theta}\frac{\partial}{\partial \tau} \int\limits_{-
\infty}^{\xi}
\cos{ \theta}{d \xi}=- \mu \frac{\partial^2 \theta}{\partial
\xi^2} .
\]
Differentiating Eq.(\ref{63}) with respect to $ \xi $ and simplifying
we obtain
\[
\displaystyle{\frac{\partial }{\partial \xi} \left[
\frac{\displaystyle{
\frac{\partial
\theta}{\partial \tau}+ \mu\frac{\partial^3 \theta}{\partial \xi^3}
}}
{\displaystyle{\frac{
\partial \theta}{\partial \xi}}}\right]=- \mu\frac{\partial^2
\theta}{\partial \xi^2}\frac{\partial \theta}{\partial \xi} .}
\]
This can be integrated   with respect to  $ \xi $   to     give,
\[
\frac{\partial \theta}{\partial \tau}+\mu\frac{\partial^3
\theta}{\partial \xi^3}= -\mu
\int\limits_{-\infty}^{\xi}\frac{\partial^2 \theta}{\partial
\xi^2}\frac{\partial \theta}{\partial \xi}{d \xi} .
\]
Multiplying throughout by $ \ds \frac{\partial \theta}{\partial \tau}
$ we get,
\[
\frac{\partial \theta}{\partial \tau}+ \mu \frac{\partial^3
\theta}{\partial \xi^3}=- \mu \frac{1}{2}
\left(\frac{\partial \theta}{\partial \xi}\right)^2\frac{\partial
\theta}{\partial \xi}.
\]
Putting $\ds f=\frac{\partial \theta}{\partial \xi} $, the above
equation becomes
\[
\frac{\partial f}{\partial \tau}+ \frac{3}{2} \mu f^2 \frac{\partial
f}{\partial \xi} +\mu \frac{\partial^3  f}{\partial \xi^3}=0.
\]
This equation is the modified Kortweg-de Vries (mKdV) equation. In the
case of steady propagation of the wave this equation can be
integrated to give a  soliton solution 
\[
f( \zeta) =2a\;\mbox{sech}\left(a\:\zeta \right)
\]
With $ \zeta =\xi-  \lambda \tau$,     $\lambda  = \mbox{constant}$,
$\ds a^2=\frac{ \lambda}{ \mu} $ if and only if $ \lambda > 0 $ $(
\mu> 0 $).


\begin{figure}[t]
\begin{picture}(10,10)
\put(20,5){\special{em: graph ris-1a.pcx}} 
\end{picture}

\vspace{5cm}

\centerline{Fig.~1. Shows the variation of $f(\xi) $ with respect to
$\xi$.\qquad\qquad\qquad\qquad\qquad}

\vspace{-1mm}

\end{figure}

Fig.~1 shows the variation of $ f(\xi) $ with respect to $ \xi $.
Figs.~2 and 3
show the variation of $ f(\zeta) $ with respect to $ \xi$ and $ \tau$
for different values of $\lambda $.
Since $\ds f=\frac{\partial \theta}{\partial \xi},\ \theta $ is defined as
\[
      \theta=  \arccos(1-2 \mbox{sech}^2 a \zeta).
\]
It is seen that $ \theta $ increases from $ 0$ to $ 2 \pi $ or
decreases from $ 0$ to $ -2 \pi $
according  as $a> 0$ or $a < o$ as $\zeta$  goes  from  $-\infty$
 to $ \infty $, since $ \theta $
is given by $ \ds \theta= \int\limits_{-\infty}^{\zeta}f{d \zeta} $.

\strut\hfill

\section{Conclusion}

Starting from the basic equations describing the propagation of
electromagnetic waves through the cold plasma we have showed that the
system of equations can be reduced to  mKdV eqation.

\begin{picture}(3,3)
\put(10,10){\special{em: graph ris-2a.pcx}}
\end{picture}

\vspace{5cm}

\centerline{Fig.~2. Show the variation of $ f(\zeta)$ with respect to
$\xi$ and $\tau$ for $\lambda = 21$.}

\strut\hfill



\begin{picture}(3,3)
\put(10,0){\special{em: graph ris-3a.pcx}}
\end{picture}

\vspace{5.5cm}

\centerline{Fig. 3. Show the variation of $ f(\zeta)$ with respect to
$\xi$ and $\tau$ for $\lambda = 30$. }





\strut\hfill

\strut\hfill

\section*{Acknowledgement}

The authors are thankful to Prof K. Babu Joseph for many valuable
discussions. One of us VCK is thankful to D.S.T, Government of India
for financial support under a research project and also wishes to
thank the Director and the IUCAA, Pune for warm hospitality and
library facilities extended to him.

\label{bindu-lp}

\end{document}